\documentclass[aoas,preprint]{imsart}

\usepackage{amsthm,amsmath,amsfonts,amssymb}
\RequirePackage[colorlinks,citecolor=blue,urlcolor=blue]{hyperref}
\usepackage[round]{natbib}
\usepackage[english]{babel}
\usepackage{latexsym}
\usepackage{subfigure}
\usepackage{epsfig}
\usepackage{color,soul}
\usepackage{moreverb}
\usepackage{mathtools}
\usepackage{enumerate,linegoal}
\usepackage{calc}
\usepackage{latexsym}
\usepackage{bm}
\usepackage{psfrag}
\usepackage{graphicx}
\usepackage{epstopdf}
\usepackage{booktabs}

\startlocaldefs
\newcommand{\expect}{\operatorname{E}}
\newcommand{\var}{\operatorname{var}}
\newcommand{\cov}{\operatorname{cov}}

\newtheorem{remark}{Remark}[section]
\newcommand{\rd}{\mathrm{d}}
\newcommand{\rds}{\,\rd}
\renewcommand{\eqref}[1]{(\ref{#1})}

\def\bSig\mathbf{\Sigma}

\def\eqd{\,{\buildrel d \over =}\,}

\newtheorem{definition}{Definition}
\endlocaldefs

\begin{document}

\begin{frontmatter}

\title{Mixed-effects models using the normal and the Laplace distributions: A $\mathbf{2 \times 2}$ convolution scheme for applied research}
\runtitle{Normal-Laplace convolutions}

\begin{aug}
\author{\fnms{Marco} \snm{Geraci}\corref{}\thanksref{t1}\ead[label=e1]{geraci@mailbox.sc.edu}}

\thankstext{t1}{Corresponding author: Marco Geraci, Department of Epidemiology and Biostatistics, Arnold School of Public Health, University of South Carolina, 915 Greene Street, Columbia SC 29209, USA. \printead{e1}}

\runauthor{M. Geraci}

\affiliation{University of South Carolina\thanksmark{t1}}

\end{aug}

\begin{abstract}
\quad In statistical applications, the normal and the Laplace distributions are often contrasted: the former as a standard tool of analysis, the latter as its robust counterpart. I discuss the convolutions of these two popular distributions and their applications in research. I consider four models within a simple $2\times 2$ scheme which is of practical interest in the analysis of clustered (e.g., longitudinal) data. In my view, these models, some of which are less known than others by the majority of applied researchers, constitute a `family' of sensible alternatives when modelling issues arise. In three examples, I revisit data published recently in the epidemiological and clinical literature as well as a classic biological dataset.
\end{abstract}

\begin{keyword}[class=MSC]
\kwd[Primary ]{62F99}
\kwd[; secondary ]{62J05}
\end{keyword}

\begin{keyword}
\kwd{Crohn's disease}
\kwd{linear quantile mixed models}
\kwd{meta-analysis}
\kwd{multilevel designs}
\kwd{random effects}
\end{keyword}

\end{frontmatter}

\section{Introduction}\label{sec:1}

The normal (or Gaussian) distribution historically has played a prominent role not only as limiting distribution of a number of sample statistics, but also for modelling data obtained in empirical studies. Its probability density is given by
\begin{equation}\label{eq:1}
f_{N}(t) = \frac{1}{\sqrt{2\pi}\sigma}  \exp \left\{ -\frac{1}{2}\left(\frac{t-\mu}{\sigma}\right)^2 \right\},
\end{equation}
for $- \infty < t < \infty$. The Laplace (or double exponential) distribution, like the normal, has a long history in Statistics. However, despite being of potentially great value in applied research, it has never received the same attention. Its density is given by
\begin{equation}\label{eq:2}
f_{L}(t) = \frac{1}{\sqrt{2}\sigma} \exp \left\{ -\sqrt{2}\left|\frac{t-\mu}{\sigma}\right| \right\}.
\end{equation}
Throughout this paper, these distributions will be denoted by $\mathcal{N}(\mu, \sigma)$ and $\mathcal{L}(\mu, \sigma)$, respectively.

\begin{figure}
\centerline{
\includegraphics[scale = 0.35]{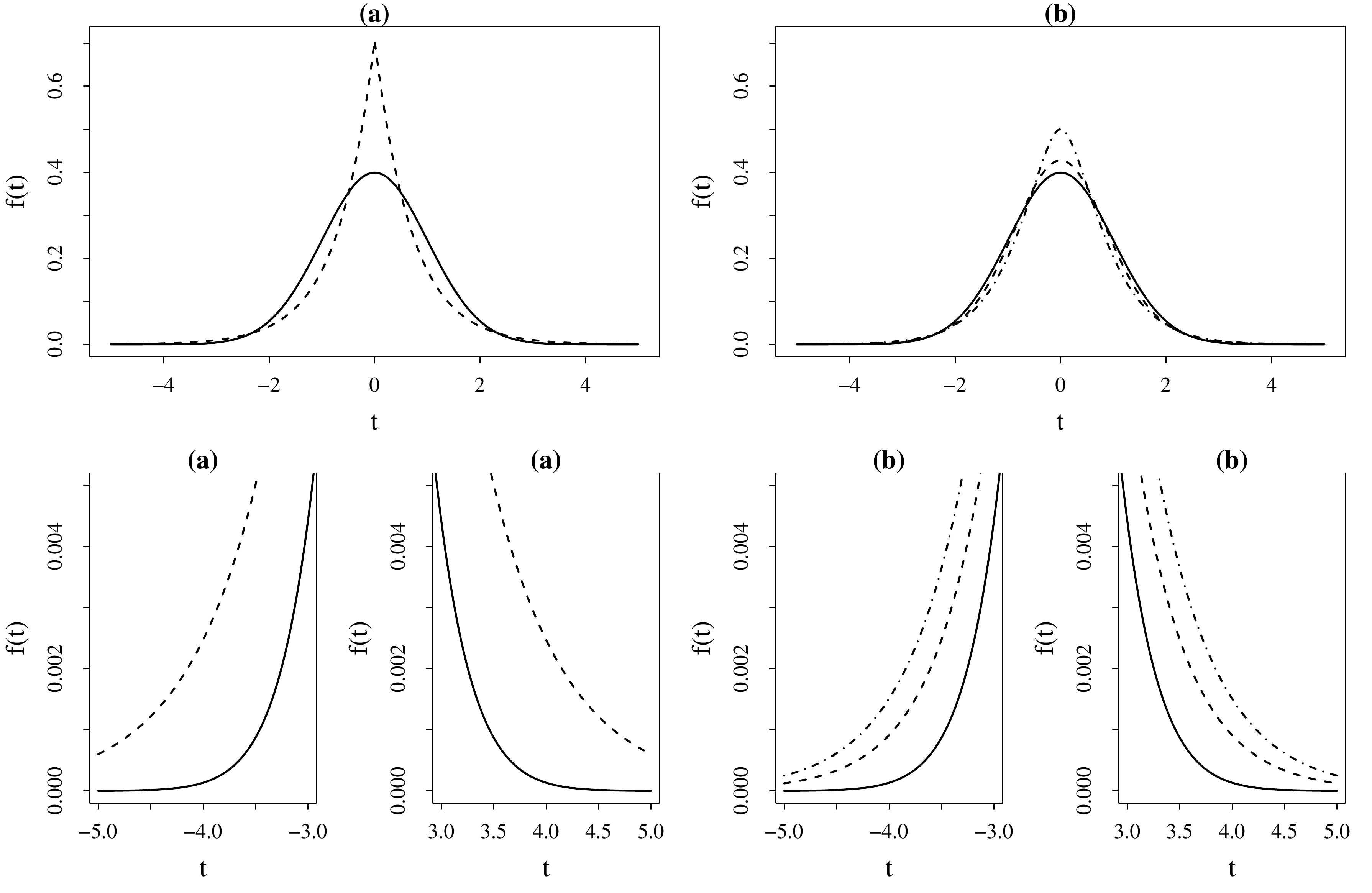}}
\caption{(a) Left: The normal (solid line) and double exponential (dashed line) densities. The location parameter is set to $0$ and the variance is set to $1$. (b) Right: The normal-normal (solid line), normal-Laplace (dashed line), and Laplace-Laplace (dot-dash line) densities. The location parameter is set to $0$ and the variance is set to $1$.}
\label{fig:1}
\end{figure}

In (\ref{eq:1}) and (\ref{eq:2}), $\mu$ and $\sigma$, where $-\infty < \mu < \infty$ and $\sigma > 0$, represent a location and a scale parameters, respectively. These two densities are shown in the left-hand side plots of Figure~\ref{fig:1}. The normal and Laplace distributions are both symmetric about $\mu$ and have variance equal to $\sigma^2$. As compared to the normal one, the Laplace density has a more pronounced peak (a characteristic technically defined \textit{leptokurtosis}) and fatter tails. Interestingly, the Laplace distribution can be represented as a scale mixture of normal distributions. Let $T \sim \mathcal{L}(\mu, \sigma)$, then \citep{kotz_2001}
\begin{equation}
\nonumber T \eqd \mu + \sigma \sqrt{E}Z,
\end{equation}
where $E$ and $Z$ are independent standard exponential and normal variables, respectively.  That is, the Laplace distribution emerges from heterogeneous normal sub-populations.

Both laws were proposed by Pierre-Simon Laplace: the double exponential in 1774 and the normal in 1778 (for an historical account, see \citeauthor{wilson_1923}, \citeyear{wilson_1923}). At Laplace's time, the problem to be solved was that of estimating $\mu$ according to the linear model
\begin{equation}
\nonumber T = \mu + \sigma\,\varepsilon,
\end{equation}
where $\varepsilon$ denotes the error term. This problem was encountered, for example, in astronomy and geodesy, where $\mu$ represented the `true' value of a physical quantity to be estimated from experimental observations. It is well known that, under the Gaussian error law (\ref{eq:1}), the maximum likelihood estimate of $\mu$ is the sample mean but, under the double exponential error law (\ref{eq:2}), it is the sample median. The former is the minimiser of the least squares (LS) estimator, while the latter is the minimiser of the least absolute deviations (LAD) estimator.

The robustness of the LAD estimator in presence of large errors was known to Laplace himself. However, given the superior analytical tractability of the LS estimator (and therefore of the normal distribution), the mean regression model
\begin{eqnarray*}
T = x^{\top}\beta + \sigma\,\varepsilon, & \; \varepsilon \sim \mathcal{N}(0,1),
\end{eqnarray*}
quickly became the `standard' tool to study the association between the location parameter of $T$ (the response variable) and other variables of interest, $x$ (the covariates).

In the past few years, theoretical developments related to least absolute error regression \citep{bassett_koenker,koenker_bassett} have led to a renewed interest in the Laplace distribution and its asymmetric extension \citep{yu_zhang} as pseudo-likelihood for quantile regression models of which median regression is a special case \citep[see, among others,][]{yu_moyeed,yu_etal,geraci_bottai_2007}. In parallel, computational advances based on interior point algorithms have made LAD estimation a serious competitor of LS methods \citep{portnoy_koenker,koenker_ng}. Another reason for the `comeback' of the double exponential is related to its robustness properties which makes this distribution and distributions alike desirable in many applied research areas \citep{kozu_nada_2010}.

In statistical applications, the interest is often in processes where the source of randomness can be attributed to more than one `error' (a hierarchy of errors is also established). For instance, this is the case of longitudinal studies where part of the variation is attributed to an individual source of heterogeneity (often called `random effect'), say $\varepsilon_{1}$, independently from the noise, $\varepsilon_{2}$, i.e.
\begin{equation}
\nonumber T = \mu + \sigma_{1}\,\varepsilon_{1} + \sigma_{2}\,\varepsilon_{2},
\end{equation}
where the distributions of $\varepsilon_{1}$ and $\varepsilon_{2}$ are often assumed to be symmetric about zero. It will be shown later that this model can be extended to include covariates associated with the parameter $\mu$ and the random effect $\varepsilon_{1}$. For now, it suffices to notice that the linear combination of random errors leads to the study of convolutions. So let us define a convolution \citep{mood_etal}.
\begin{definition}
If $U$ and $V$ are two independent, absolutely continuous random variables with density functions $f_{U}$ and $f_{V}$, respectively, and $T = U + V$, then
\[
f_{T}(t) = f_{U+V}(t) = \int_{-\infty}^{\infty} f_{V}(t - u)f_{U}(u)\,du = \int_{-\infty}^{\infty} f_{U}(t - v)f_{V}(v)\rds v
\]
is the convolution of $f_{U}$ and $f_{V}$.
\end{definition}

In Section \ref{sec:2}, I consider convolutions based on the normal and the Laplace distributions within a simple and practical $2 \times 2$ scheme. In Section \ref{sec:3}, I discuss inference when data are clustered, along with the implementation of estimation procedures using existing R \citep{R} software (further technical details are provided in Appendix, along with a simulation study). In Section \ref{sec:4}, I show some applications and, in Section \ref{sec:5}, conclude with final remarks.

\section{Convolutions}\label{sec:2}
Let $Y$ be a real-valued random variable with absolutely continuous distribution function $F(y) = \Pr\left\{Y \leq y \right\}$ and density $f(y)\equiv F'(y)$. The variable $Y$ is observable and represents the focus of the analysis in specific applications (e.g., as the response variable in regression models). I consider four cases in which $Y$ results from one of the four convolutions reported in Table~\ref{tab:1}. The letters $\nu$ and $\lambda$ are used to denote normal and Laplace variates with densities (\ref{eq:1}) and (\ref{eq:2}), respectively. The subscripts 1 and 2 indicate, respectively, which of the two random variables plays the role of a random effect and which one is considered to be the noise. Here, the former may in general be associated with a vector of covariates and may represent an inferential quantity of interest; the latter is treated as a nuisance. Moreover, I assume independence between the components of the convolution throughout the paper.

\begin{table}[t!]
\caption{$2 \times 2$ convolution scheme for independent Gaussian ($\nu$) and Laplacian ($\lambda$) random variables.}
\centering
\begin{tabular}{lccp{3cm}p{3cm}p{3cm}}
  \toprule
   & Normal & Laplace \\
  \hline
  Normal & $\nu_{1} + \nu_{2}$ (NN) &  $\nu_{1} + \lambda_{2}$ (NL)\\
  Laplace & $\lambda_{1} + \nu_{2}$ (LN) &  $\lambda_{1} + \lambda_{2}$ (LL)\\
  \hline
\end{tabular}\label{tab:1}
\end{table}

A few remarks about notation are needed. The shorthand $\mathrm{diag}(t)$ or $\mathrm{diag}(t_{1}, \ldots, t_{n})$, where $t = (t_{1},\ldots,t_{n})^{\top}$ is a $n \times 1$ vector, is used to denote the $n \times n$ diagonal matrix whose diagonal elements are the corresponding elements of $t$. The standard normal density and cumulative distribution functions will be denoted by $\phi$ and $\Phi$, respectively.

\subsection{Normal-normal (NN) convolution}\label{sec:2.1}
The first convolution
\begin{equation}\label{eq:3}
Y = \nu_{1} + \nu_{2},
\end{equation}
where $\nu_{1} \sim \mathcal{N}(0, \sigma_{1})$ and $\nu_{2} \sim \mathcal{N}(0, \sigma_{2})$, represents, in some respects, the simplest case among the four combinations defined in Table~\ref{tab:1}. Standard theory of normal distributions leads to
\begin{equation}\label{eq:4}
f_{NN}(y) = \frac{1}{\psi}\; \phi\left(\frac{y}{\psi}\right),
\end{equation}
where $\psi^2 \equiv \var(Y) =  \sigma_{1}^2 + \sigma_{2}^2$.

Model (\ref{eq:3}) can be generalised to the regression model
\begin{equation}\label{eq:5}
Y = x^{\top}\beta + z^{\top}\nu_{1} + \nu_{2},
\end{equation}
where $x$ and $z$ are, respectively, $p \times 1$ and $q \times 1$ vectors of covariates, and $\beta$ is a $p\times 1$ dimensional vector of regression coefficients. If $q > 1$, then I assume $\nu_1 \sim \mathcal{N}_{q}(0, \Sigma_{1})$, that is, a multivariate normal distribution with $q \times q$ variance-covariance matrix $\Sigma_{1}$. It follows that
\begin{equation}\label{eq:6}
g_{NN}(y) = \frac{1}{\psi}\; \phi\left(\frac{y - x^{\top}\beta}{\psi}\right),
\end{equation}
where $\psi^2 \equiv \var(Y) = z^{\top} \Sigma_{1} z + \sigma_{2}^2$.

Model~(\ref{eq:5}) is known as a linear mixed effects (LME) model or, simply, as a mixed model \citep{pinheiro_bates,demidenko_2013}. There is a vast number of applications of LME models, especially for the analysis of clustered data in the social, life and physical sciences.

\subsection{Normal-Laplace (NL) convolution}\label{sec:2.2}

The second convolution consists of a normal and a Laplace components, that is
\begin{equation}\label{eq:7}
Y = \nu_{1} + \lambda_{2},
\end{equation}
where $\nu_{1} \sim \mathcal{N}(0, \sigma_{1})$ and $\lambda_{2} \sim \mathcal{L}(0, \sigma_{2})$. The resulting density is given by \citep{reed_2006}
\begin{equation}\label{eq:8}
f_{NL}(y) = \frac{1}{\sqrt{2}\sigma_{2}}\;\phi\left(y/\sigma_{1}\right)\left\{R\left(\kappa - y/\sigma_{1}\right) + R\left(\kappa + y/\sigma_{1}\right)\right\},
\end{equation}
where $\kappa = \sqrt{2}\sigma_{1}/\sigma_{2}$ and $R$ is the Mills ratio
\[
R(t) = \frac{1 - \Phi(t)}{\phi(t)}.
\]
The above distribution arises from a Brownian motion whose starting value is normally distributed and whose stopping hazard rate is constant. An extension of (\ref{eq:8}) to skewed forms can be obtained by letting $\lambda_{2}$ follow an asymmetric Laplace distribution \citep{reed_2006}. Applications of the NL convolution can be found in finance \citep{reed_2007,meintanis_2010}. See also the double Pareto-lognormal distribution, associated with $\exp(Y)$, which has applications in modeling size distributions \citep{reed_jorgensen}.

As in the previous section, I consider a generalisation of model (\ref{eq:7}) to the regression case
\begin{equation}\label{eq:9}
Y = x^{\top}\beta + z^{\top}\nu_{1} + \lambda_{2}.
\end{equation}
If $q > 1$, then I assume $\nu_1 \sim \mathcal{N}_{q}(0, \Sigma_{1})$. It follows that $z^{\top}\nu_{1}$ is normal with mean zero and variance $z^{\top} \Sigma_{1}z$. This leads to
\begin{equation}\label{eq:10}
g_{NL}(y) = \frac{1}{\sqrt{2}\sigma_{2}}\;\phi\left(\frac{y - x^{\top}\beta}{\sigma_{1}}\right)\left\{R\left(\kappa - \frac{y - x^{\top}\beta}{\sigma_{1}}\right) + R\left(\kappa + \frac{y - x^{\top}\beta}{\sigma_{1}}\right)\right\},
\end{equation}
where $\sigma_{1} \equiv \sqrt{z^{\top} \Sigma_{1}z}$ and, as defined above, $\kappa = \sqrt{2}\sigma_{1}/\sigma_{2}$. It is easy to verify that $\var(Y) =  \sigma_{1}^2 + \sigma_{2}^2$.

Model~(\ref{eq:9}) is a median regression model with normal random effects, a special case of the linear quantile mixed models (LQMMs) discussed by \cite{geraci_bottai_2007,geraci_bottai_2014}. LQMMs have been used in a wide range of research areas, including marine biology \citep{muir_etal_2015,duffy_etal_2015,barneche_2106}, environmental science \citep{fornaroli_etal_2015}, cardiovascular disease \citep{degerud_2014,blankenberg_2016}, physical activity \citep{ng,beets}, and ophthalmology \citep{patel_etal_2015,patel_etal_2016}.

\subsection{Laplace-normal (LN) convolution}\label{sec:2.3}
The Laplace-normal convolution is given by
\begin{equation}\label{eq:11}
Y = \lambda_{1} + \nu_{2},
\end{equation}
where $\lambda_{1} \sim \mathcal{L}(0, \sigma_{1})$ and $\nu_{2} \sim \mathcal{N}(0, \sigma_{2})$. The LN appears in robust meta-analysis \cite[p.266]{demidenko_2013}.

The LN convolution in (\ref{eq:11}), clearly, is the same as the NL convolution in (\ref{eq:7}) (so I omit writing its density). However, note that now the Laplace component is associated with the random effect, not with the error term; therefore, the two scale parameters $\sigma_{1}$ and $\sigma_{2}$ will appear swapped. The distinction becomes clear when considering the regression model
\begin{equation}\label{eq:12}
Y = x^{\top}\beta + z^{\top}\lambda_{1} + \nu_{2}.
\end{equation}
By analogy with the NL convolution, I assume that, for $q > 1$, $\lambda_{1}$ has a $q$-dimensional multivariate Laplace distribution \cite[p.235]{kotz_2001}.
\begin{definition}
An $n$-dimensional random variable $T$ is said to follow a zero-centred multivariate Laplace distribution with parameter $\Sigma$, $T \sim \mathcal{L}_{q}(0,\Sigma)$, if its density is given by
\[
f_{L}(t) = 2(2\pi)^{-n/2}\;\left|\Sigma\right|^{-1/2}\left(t^{\top} \Sigma^{-1} t/2\right)^{\omega/2}K_{\omega}\left(\sqrt{2t^{\top} \Sigma^{-1} t}\right),
\]
where $\Sigma$ is an $n \times n$ nonnegative definite symmetric matrix, $\omega = (2-n)/2$ and $K_{\omega}$ is the modified Bessel function of the third kind.
\end{definition}

\begin{remark}
If $T \sim \mathcal{L}_{q}(0,\Sigma)$, then $\cov(t) = \Sigma$ \cite[p.249]{kotz_2001}. For a diagonal matrix $\Sigma = \mathrm{diag}(\varsigma_{1}, \ldots, \varsigma_{q})$, the coordinates of the multivariate Laplace are uncorrelated, but not independent. Therefore, the joint distribution of $n$ independent univariate Laplace variates does not have the properties of the multivariate Laplace with diagonal variance-covariance matrix.
\end{remark}

For $n=1$, the multivariate density defined above reduces to the univariate density (\ref{eq:2}) with $\sigma = \Sigma^{1/2}$. Moreover, a linear combination of the coordinates of the multivariate Laplace is still a Laplace \cite[p.255]{kotz_2001}. Indeed, if we assume $\lambda_{1} \sim \mathcal{L}_{q}(0,\Sigma_{1})$, then $z^{\top}\lambda_{1} \sim \mathcal{L}(0,\sigma_{1})$, where $\sigma_{1} = \sqrt{z^{\top} \Sigma_{1} z}$. Thus, the density of $Y$ in Equation (\ref{eq:12}) is given by
\begin{equation}\label{eq:13}
g_{LN}(y) = \frac{1}{\sqrt{2}\sigma_{1}}\;\phi\left(\frac{y - x^{\top}\beta}{\sigma_{2}}\right)\left\{R\left(\kappa - \frac{y - x^{\top}\beta}{\sigma_{2}}\right) + R\left(\kappa + \frac{y - x^{\top}\beta}{\sigma_{2}}\right)\right\},
\end{equation}
where $\kappa = \sqrt{2}\sigma_{2}/\sigma_{1}$. Again, it is easy to verify that $\var(Y) = \sigma_{1}^2 + \sigma_{2}^2$.

\subsection{Laplace-Laplace (LL) convolution}\label{sec:2.4}
The fourth and last convolution consists of two Laplace variates, i.e.
\begin{equation}\label{eq:14}
Y = \lambda_{1} + \lambda_{2},
\end{equation}
where $\lambda_{1} \sim \mathcal{L}(0, \sigma_{1})$ and $\lambda_{2} \sim \mathcal{L}(0, \sigma_{2})$. It can be shown \cite[p.35]{kotz_2001} that the density of $Y$ is
\begin{equation}\label{eq:15}
f_{LL}(y) =
\begin{cases}
\dfrac{1}{4}s (1 + s|y|)\exp(-s|y|), & \text{if $s_{1} = s_{2} = s,$}\\
\dfrac{\kappa}{2\kappa^2 - 2} \left\{s_{1}\exp(-s_{2}|y|) - s_{2}\exp(-s_{1}|y|)\right\}, & \text{if $s_{1}/s_{2} = \kappa \neq 1,$}
\end{cases}
\end{equation}
with $s_{1} = \sigma_{1}/\sqrt{2}$ and $s_{2} = \sigma_{2}/\sqrt{2}$.

For the regression model
\begin{equation}\label{eq:16}
Y = x^{\top}\beta + z^{\top}\lambda_{1} + \lambda_{2},
\end{equation}
with $\lambda_{1} \sim \mathcal{L}_{q}(0,\Sigma_1)$, I obtain
\begin{equation}\label{eq:17}
g_{LL}(y) =
\begin{cases}
\dfrac{1}{4}s (1 + s|y - x^{\top}\beta|)\exp(-s|y - x^{\top}\beta|), & \text{if $s_{1} = s_{2} = s,$}\\
\dfrac{\kappa}{2\kappa^2 - 2} \left\{s_{1}\exp(-s_{2}|y - x^{\top}\beta|) \right.& \\
\left. \qquad - s_{2}\exp(-s_{1}|y - x^{\top}\beta|)\right\}, & \text{if $s_{1}/s_{2} = \kappa \neq 1.$}
\end{cases}
\end{equation}
with $s_{1} = \sigma_{1}/\sqrt{2}$ and $\sigma_{1} = \sqrt{z^{\top} \Sigma_{1} z}$. The variance is given by $\var(Y) =  \sigma_{1}^2 + \sigma_{2}^2$.

Model~(\ref{eq:16}) is a median regression model with `robust' random effects, another special case of LQMMs \citep{geraci_bottai_2014}.

\subsection{Some properties}\label{sec:2.5}
All the convolutions are symmetric, unimodal, twice differentiable and have continuous first and second derivatives (the NN and NL are also smooth). Also, they are log-concave since both the normal (\ref{eq:1}) and Laplace (\ref{eq:2}) densities are log-concave \citep{prekopa_1973}. The right-hand side plots of Figure~\ref{fig:1} shows that, as compared to the NN density, the NL (LN) and LL densities are leptokurtic and have more weight in the tails, with the NL density sitting between the NN and LL distributions. Thus, the presence of the Laplace term in the convolution confers different degrees of robustness to the model depending on whether one or both random terms are assumed to be Laplacian. Also, notice that the marginal regression models are location--scale-shift models, since both the location and the scale of $Y$ are functions of the covariates.

\section{Inference}\label{sec:3}

In this section, I briefly discuss inferential issues, with detailed mathematical derivations provided in Appendix.

Let $Y_{i} = (Y_{i1}, Y_{i2}, \ldots, Y_{in_{i}})^{\top}$ be a multivariate $n_{i} \times 1$ random response vector, and $x_{ij}$ and $z_{ij}$, be vectors of covariates for the $j$th observation, $j = 1, \ldots, n_{i}$, in cluster $i$, $i = 1, \ldots, M$. Each component of $Y_{i}$ can be modelled using any of the convolutions discussed in Section \ref{sec:2} by assuming
\begin{equation}
\nonumber Y_{ij} = x_{ij}^{\top}\beta + z_{ij}^{\top}\varepsilon_{1i} + \varepsilon_{2ij},
\end{equation}
where the random effect $\varepsilon_{1i}$ and the error term $\varepsilon_{2ij}$ are either Gaussian or Laplacian according to the scheme in Table \ref{tab:1}. The marginal models implied by these four convolutions have been defined in expressions (\ref{eq:6}), (\ref{eq:10}), (\ref{eq:13}), and (\ref{eq:17}). At the cluster level, I use the notation
\begin{equation}\label{eq:18}
Y_{i} = X_{i}\beta + Z_{i}\varepsilon_{1i} + \varepsilon_{2i},
\end{equation}
where $X_{i}$ and $Z_{i}$ are, respectively, $n_{i} \times p$ and $n_{i} \times q$ design matrices. I assume that the vector of random effects $\varepsilon_{1i}$ has variance-covariance matrix $\Sigma_{1}$ for all $i = 1, \ldots, M$ and that the $Y_{i}$'s are independent of one another. The structure of $\Sigma_1$ is, for the moment, left unspecified. Also, I assume that $\mathrm{cov}(\varepsilon_{2i})$ is a multiple of the identity matrix, although this assumption can be easily relaxed (see Section \ref{sec:3.4}).

There are several approaches to mixed effects model estimation \citep[see, for example,][]{pinheiro_bates,demidenko_2013}, each approach having its own advantages and disadvantages. One approach is to work with the marginal likelihood of $Y_{i}$. Although independence between clusters can still be assumed, in general the $Y_{ij}$'s will be correlated within the same cluster. Therefore, parameter estimation based on the marginal likelihood requires knowing the joint distribution of $Y_{i1}, Y_{i2}, \ldots, Y_{in_{i}}$. Under the NN convolution, $Y_{i}$ is known to be multivariate normal. It is beyond the scope of this paper to derive the multivariate distribution of $Y_{i}$ for the NL, LN and LL convolutions.

Likelihood-based estimation of location and scale parameters using the NN model has been largely studied. Therefore, I will focus on the NL, LN, and LL models. Since an important aspect in applied research is the availability of software to perform data analysis, here I consider two methods which can be applied using existing software. The first method is based on numerical integration and applies to specific NL and LL models, while the other method is based on a Monte Carlo Expectation-Maximisation (MCEM) algorithm and applies to NL, LN and LL models.

\subsection{Numerical integration}\label{sec:3.1}

Let the $i$th contribution to the marginal log-likelihood be
\begin{equation}
\nonumber \ell(\beta, \Sigma_1, \sigma_2; y_{i}) = \log \int_{\mathbb{R}^{q}} g\left(y_{i} - X_{i}\beta - Z_{i}u_{i}\right) h(u_{i}) \rds u_{i},
\end{equation}
where $g$ denotes the density of the error term conditional on the random effect $u_i$ and $h$ denotes the density of the random effect. One can work with the numerically integrated likelihood
\begin{equation}\label{eq:19}
\tilde{\ell}(\beta, \Sigma_1, \sigma_2; y_{i}) = \log \sum_{k_{1}}^{K}\cdots\sum_{k_{q}}^{K} g\left(y_{i} - X_{i}\beta - Z_{i}\left(\Sigma_{1}\right)^{1/2}v_{k_{1},\ldots,k_{q}}\right) h(v_{k_{1},\ldots,k_{q}}),
\end{equation}
with nodes $v_{k_{1},\ldots,k_{q}} = (v_{k_{1}}, \ldots, v_{k_{q}})^{\top}$ and weights $h(v_{k_{1},\ldots,k_{q}})$, $k_{l} = 1,\ldots,K$, $l = 1, \ldots, q$, as an approximation to the marginal log-likelihood.

The maximisation of the approximate log-likelihood (\ref{eq:19}) can be time-consuming depending on the dimension of the quadrature $q$, the required accuracy of the approximation controlled by the number of nodes $K$, and, of course, the distribution $h(u)$. If $\Sigma_1$ is a diagonal matrix, then $h(v_{k_{1},\ldots,k_{q}})=\prod_{l=1}^{q} h(v_{k_{l}})$. This greatly simplifies calculations since the $q$-dimensional integral can be carried out with $q$ successive applications of one-dimensional quadrature rules. In the multivariate normal case, a non-diagonal covariance matrix can be rescaled to a diagonal one and the joint density factorises into $q$ normal variates. However, this is not the case for the multivariate Laplace, at least not for the one defined in Section~\ref{sec:2.3}. \cite{geraci_bottai_2014} considered a steepest-descent approach combined with Gauss-Hermite and Gauss-Laguerre quadrature for, respectively, the NL and LL likelihoods. Standard errors were obtained by bootstrapping the clusters (block bootstrap).

Since \citeauthor{geraci_2014}'s (\citeyear{geraci_2014}) algorithms, which are implemented in the R package \texttt{lqmm}, can be applied to selected models only (namely, NL models with correlated or uncorrelated random effects and LL models with uncorrelated random effects), in the next section I develop an alternative, more general approach based on the EM algorithm.

\subsection{EM estimation}\label{sec:3.2}

Rather than working with the Laplace distribution directly, I consider its representation as a scale mixture of normal distributions. As noted before, if $T \sim \mathcal{L}(0, \sigma)$, then $T \eqd \sigma\sqrt{W}V$, where $W$ and $V$ are, respectively, independent standard exponential and normal variates. This equivalence has been used in EM estimation of regression quantiles \citep[see, for example,][]{lum_gelfand,tian_etal_2013}. Similarly, in the multivariate case, if $T \sim \mathcal{L}_{q}(0,\Sigma_1)$, then $T \eqd \sqrt{W}V$, where $W$ is, again, standard exponential and $V \sim \mathcal{N}_{q}(0, \Sigma_{1})$. As shown in Appendix, the normal components in the scale mixture representation of the NL, LN, and LL models can be easily convolved (conditionally on $W$) and the resulting log-likelihood for the $i$th cluster becomes
\begin{equation}
\nonumber \ell\left(\beta, \Sigma_1, \sigma_2; y_{i},w_{i}\right) = \log g\left(y_{i}|w_{i}\right) + \log h(w_{i}),
\end{equation}
where $g$ is multivariate normal and $h$ is standard exponential.

The proposed EM algorithm starts from the likelihood of the complete data $(y_{i},w_{i})$, where $w_{i}$ represents the unobservable data. In the E-step, the expected value of the complete log-likelihood is approximated using a Monte Carlo expectation. As shown in expression~\eqref{eq:A.6} in Appendix, the M-step reduces to the maximum likelihood estimation of a linear mixed model with prior weights which can be carried out using fitting routines from existing software (e.g., \texttt{nlme} or \texttt{lme4} in R).

\subsection{Modelling and estimation of $\Sigma_1$}\label{sec:3.3}
There are different possible structures for $\Sigma_1$. The simplest is a multiple of the identity matrix, with constant diagonal elements and zero off-diagonal elements. Other structures include, for example, diagonal (variance components), compound symmetric (constant diagonal and constant off-diagonal elements), and the more general symmetric positive-definite matrix. These are all available in the \texttt{nlme} \citep{pinheiro_2014}, \texttt{lme4} \citep{bates_2015} and \texttt{lqmm} \citep{geraci_2014} packages, as well as in SAS procedures for mixed effects models.

The variance-covariance matrix of the random effects, whether normal or Laplace, must be nonnegative definite. However, it is possible that, during MLE, the estimate $\hat{\Sigma}_1$ may be singular or veer off into the space of negative definite matrices. This problem does not occur in EM estimation if the starting matrix is nonnegative definite. However, the monotonicity property is lost when a Monte Carlo error is introduced at the E-step \citep{mclachlan_2008}. There are at least three approaches one can consider \citep[p.88]{demidenko_2013}: (i) allow $\hat{\Sigma}_1$ to be negative definite during estimation and, if negative definite at convergence, replace it with a nonnegative definite matrix after the algorithm has converged; (ii) constrained optimisation; (iii) matrix reparameterisation \citep{pinheiro_bates}. As discussed in Appendix, I follow the latter approach.

\subsection{Residual heteroscedasticity and correlation}\label{sec:3.4}

The development of the EM algorithm discussed above is based on the assumption that the within-group errors are independent with common scale parameter $\sigma_{2}$. As briefly outlined in Section~\ref{sec:A.6} in Appendix, it is easy to extend the NL, LN, and LL models to the case of heteroscedastic and correlated errors. Commonly available mixed effects software provide capabilities for estimating residual variance and correlation parameters. For the sake of simplicity, I do not consider this extension any further in this paper.

\section{Examples}\label{sec:4}
\subsection{Meta-analysis}\label{sec:4.1}

\begin{figure}
\centerline{\includegraphics[scale = .9]{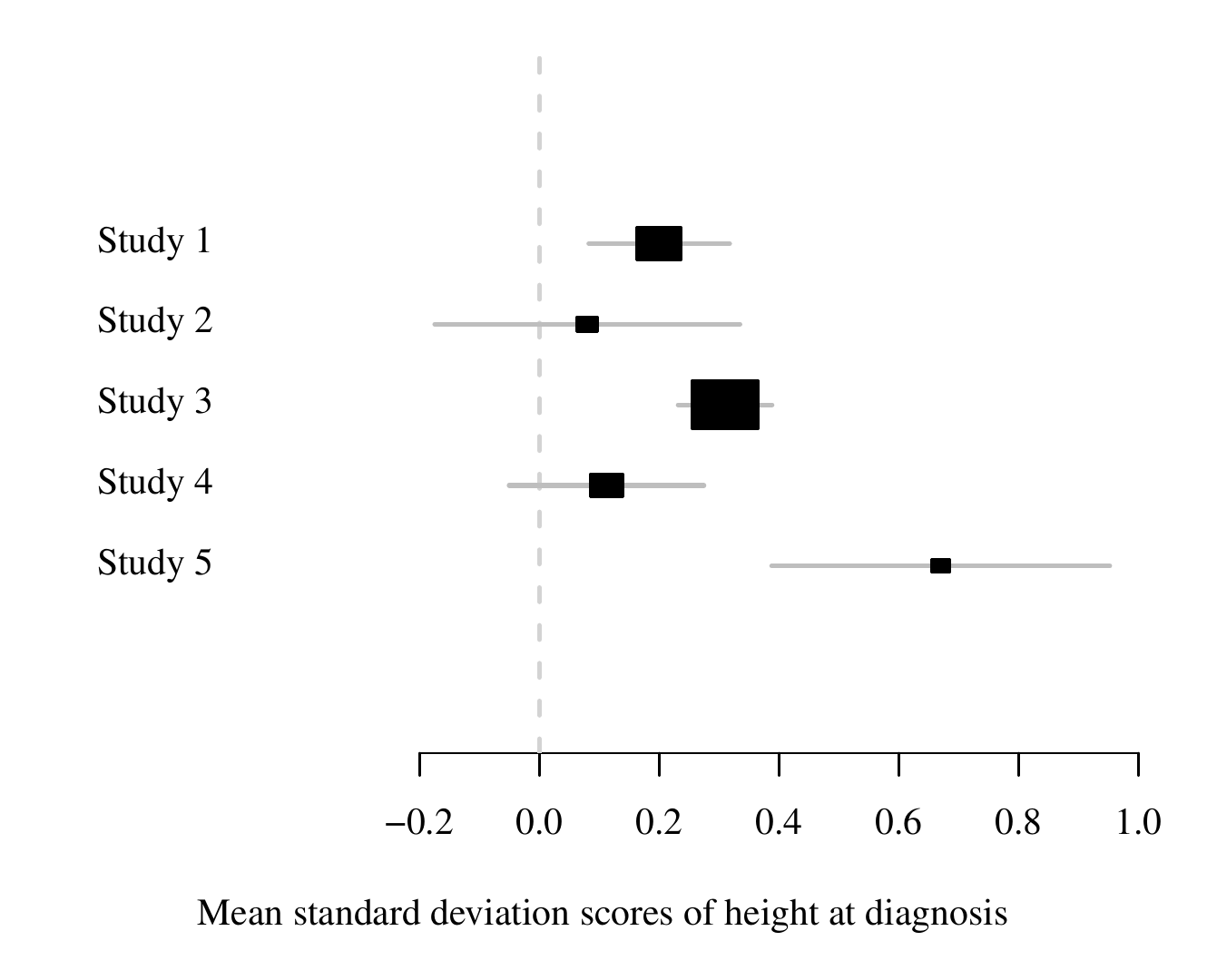}}
\caption{Forest plot for five studies on the relationship between height at diagnosis and osteosarcoma in young people. Each study is represented by a block, with area proportional to its weight, centered at the effect point estimate. Horizontal grey lines depict $95\%$ confidence intervals.}
\label{fig:2}
\end{figure}

Here, I discuss an application in meta-analysis. The data consist of mean standard deviation scores of height at diagnosis in osteosarcoma patients which had been reported in five different studies (Figure~\ref{fig:2}) and were successively meta-analysed by \cite{arora_2011}. Let $Y$ denote the study-specific effect. For these data, I considered the NN model \citep{dersimonian_laird}
\begin{eqnarray*}
Y_{i} = \mu + \nu_{1i} + \nu_{2i}, & i = 1,\ldots,5,
\end{eqnarray*}
where $\nu_{1i} \sim \mathcal{N}(0, \tau)$ and $\nu_{2i} \sim \mathcal{N}(0, \sigma_{i})$, and the LN model \citep{demidenko_2013}
\begin{eqnarray*}
Y_{i} = \mu + \lambda_{1i} + \nu_{2i}, & i = 1,\ldots,5,
\end{eqnarray*}
where $\lambda_{1i} \sim \mathcal{L}(0, \tau)$ and $\nu_{2i} \sim \mathcal{N}(0, \sigma_{i})$.

In meta-analysis, the goal is to estimate an `overall' or `pooled' effect ($\mu$) and the between-study variance or heterogeneity among study-specific effects ($\tau^2$). The sampling variances $\sigma_{i}^2$ are assumed to be known.

Estimation for the osteosarcoma data was carried out using R software developed for standard \citep{viechtbauer_2010} and robust \citep{demidenko_2013} meta-analysis. The estimates (standard errors) of $\mu$ and $\tau^2$ were, respectively, $0.260$ (0.087) and $0.029$ (0.027) for the NN model, and $0.246$ (0.073) and $0.021$ (0.033) for the LN model. The larger estimated overall effect and heterogeneity for the NN model are a consequence of the outlying effect size of study 5 (Figure~\ref{fig:2}) which skews the location $\mu$ and inflates the scale of the normal distribution. In contrast, the Laplace distribution is more robust to outliers and heavy tails. Indeed, the estimate of $\mu$ from the LN model was more precise as demonstrated by the lower standard error (as a consequence, the related test statistic has smaller $p$-value). A similar example is described by \cite{demidenko_2013}.

\subsection{Repeated measurements in clinical trials}\label{sec:4.2}

Ten Crohn's disease patients with endoscopic recurrence were followed over time \citep{sorrentino_etal_2010}. Colonoscopy was performed and surrogate markers of disease activity were collected on four occasions. One of the goals of this trial was to assess the association between fecal calprotectin (FC -- mg/kg) and endoscopic score (ES -- Rutgeerts). The data were analysed using a log-linear median regression model under the assumption of independence between measurements \citep{sorrentino_etal_2010}. Here, I take the within-patient correlation into account and analyse the data using three of the four regression models discussed in Section~\ref{sec:2}: the NN model
\begin{eqnarray*}
\log Y_{ij} = \beta_{0} + \beta_{1}x_{ij} + \nu_{1i} + \nu_{2ij}, & \; j = 1, \ldots, 4, & \; i = 1,\ldots,10,
\end{eqnarray*}
the NL model
\begin{eqnarray*}
\log Y_{ij} = \beta_{0} + \beta_{1}x_{ij} + \nu_{1i} + \lambda_{2ij}, & \; j = 1, \ldots, 4, & \; i = 1,\ldots,10,
\end{eqnarray*}
and the LL model
\begin{eqnarray*}
\log Y_{ij} = \beta_{0} + \beta_{1}x_{ij} + \lambda_{1i} + \lambda_{2ij}, & \; j = 1, \ldots, 4, & \; i = 1,\ldots,10,
\end{eqnarray*}
where $Y_{ij}$ and $x_{ij}$ denote, respectively, FC and ES measurements on patient $i$ at occasion $j$, $\nu_{1i} \sim \mathcal{N}(0, \tau)$, $\nu_{2ij} \sim \mathcal{N}(0, \sigma)$, $\lambda_{1i} \sim \mathcal{L}(0, \tau)$, and $\lambda_{2ij} \sim \mathcal{L}(0, \sigma)$. Therefore, the variance of the random effects is $\tau^2$, while the variance of the error term is $\sigma^2$.

\begin{table}[ht!]
\caption{Association between fecal calprotectin and endoscopic score in Crohn's disease patients. Estimates and standard errors (SE) of the fixed effects ($\beta$), variance of the random effects ($\tau^2$), and intra-class correlation ($\rho$) from three models. The log-likelihood ($\ell$) is reported in brackets.}
\centering
\begin{tabular}{lrrrr}
  \toprule
 & $\beta_{0}$ & $\beta_{1}$ & $\tau^2$ & $\rho$ \\
  \hline
\multicolumn{5}{l}{\textit{Normal-Normal} ($\ell = -21.8$)}\\
Estimate & 3.293 & 0.910 & 0.031 & 0.191 \\
SE & 0.113 & 0.056 & 0.133 &  \\
\multicolumn{5}{l}{\textit{Normal-Laplace} ($\ell = -22.2$)}\\
Estimate & 3.354 & 0.871 & 0.994 & 0.877 \\
SE & 0.135 & 0.051 & 0.046 &  \\
\multicolumn{5}{l}{\textit{Laplace-Laplace} ($\ell = -14.2$)}\\
Estimate & 3.269 & 0.905 & 0.293 & 0.757 \\
SE & 0.114 & 0.035 & 0.053 &  \\
\hline
\end{tabular}\label{tab:2}
\end{table}

\begin{figure}
\centerline{\includegraphics[scale = 0.5]{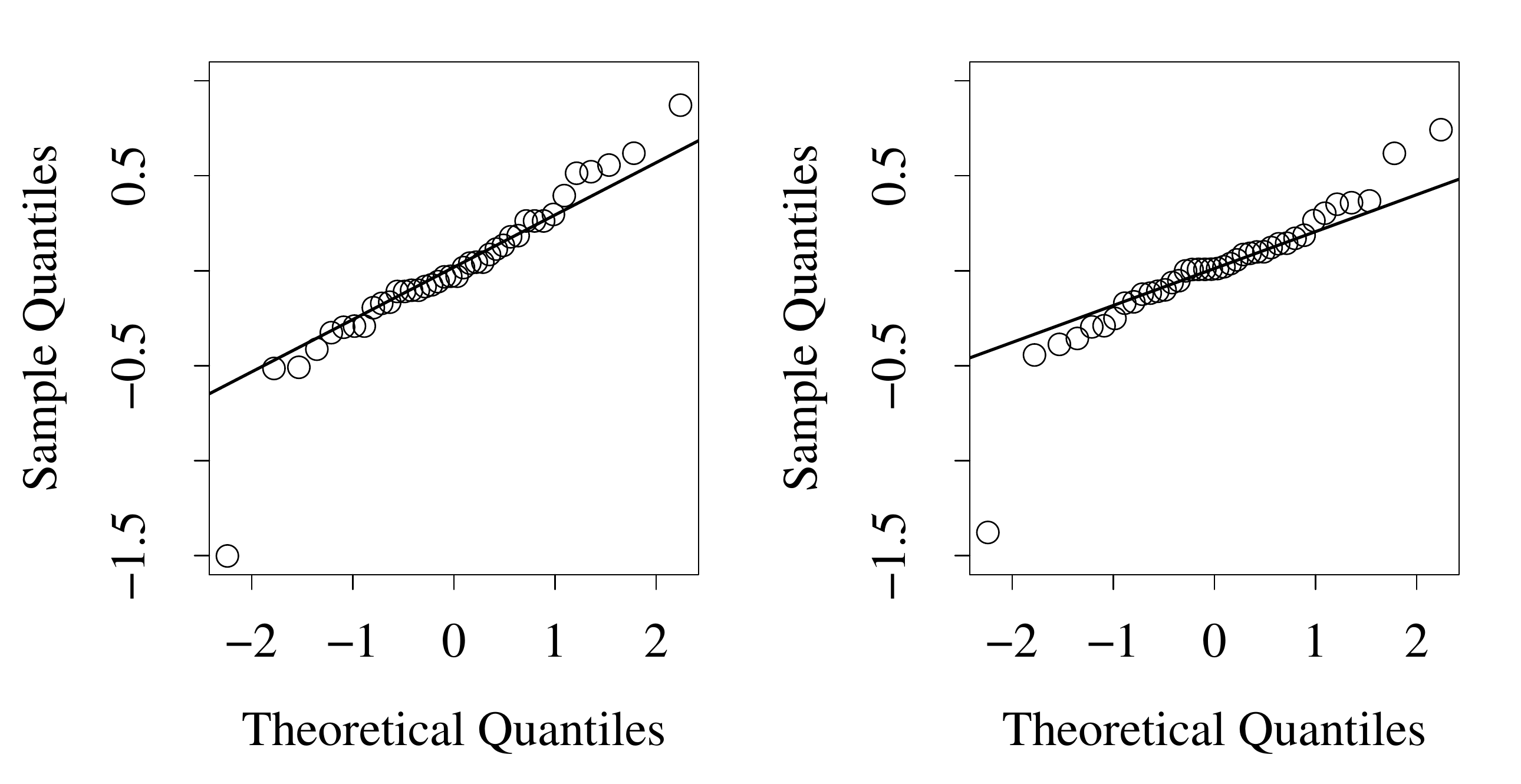}}
\caption{QQ-plot of the first-level (left plot) and second-level (right plot) residuals from the normal-normal model for the Crohn's disease data.}
\label{fig:3}
\end{figure}

In this case, the parameters of interest are the slope $\beta_1$ and the intra-class correlation $\rho = \tau^2/(\tau^2 + \sigma^2)$, which measures how much of the total variance is due to between-individual variability. Estimation was carried out using the \texttt{nlme} \citep{pinheiro_2014} and \texttt{lqmm} \citep{geraci_2014} packages. The results are shown in Table~\ref{tab:2}. The estimates of the regression coefficients $\beta$ tallied across models. However, the estimates of $\tau^2$ and $\rho$ differed substantially, with values from the NN model much lower than those from the NL and LL models. First-level residuals (i.e., predictions of the random effects plus the error term) and second-level residuals (i.e., predictions of the error term only) from the NN model are shown in Figure~\ref{fig:3}. It is apparent that $\sigma^2$ may be inflated by an unusual second-level residual, to the detriment of $\tau^2$. As a consequence, the intra-class correlation appeared to be heavily underestimated by the NN model. The NL model improved upon the estimation of the scale parameters as it is more robust to outliers in the error term. However, the LL model gave the largest value of the log-likelihood, suggesting that the goodness of the fit is further improved by using a robust distribution for the random effects as well. Note also that the standard error of the slope was smallest for the LL model.

\subsection{Growth curves}\label{sec:4.3}

\begin{figure}
\centerline{\includegraphics[scale = 0.45]{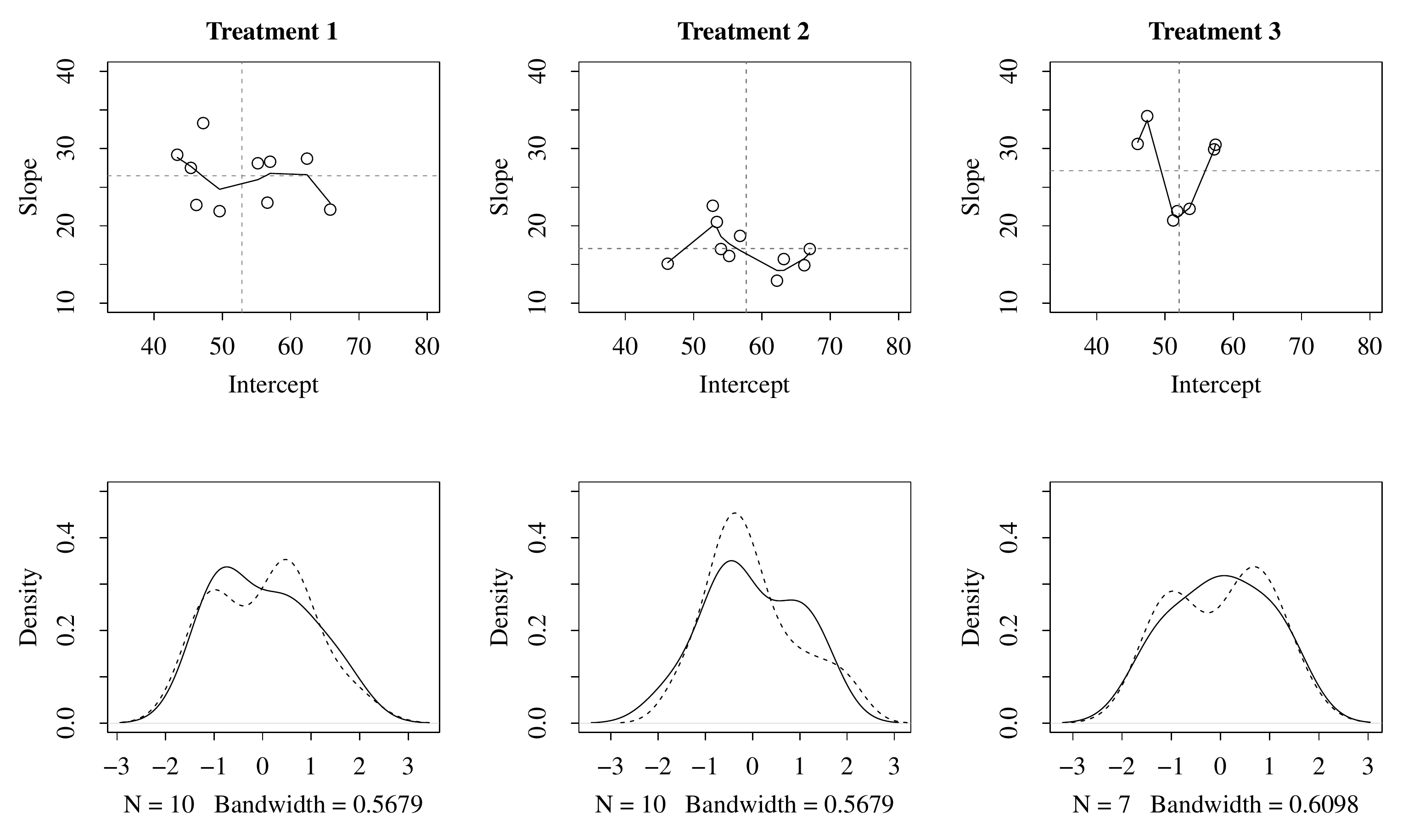}}
\caption{Ordinary least squares estimates of intercepts and slopes for individual growth curves in the rats weight gain data. The scatterplots on the top show the pairwise estimates with LOESS smoothing superimposed (dashed grey lines mark mean values). The plots on the bottom depict the estimated densities of intercepts (solid line) and slopes (dashed line) centred and scaled using their respective means and standard deviations.}
\label{fig:4}
\end{figure}

In a weight gain experiment, 30 rats were randomly assigned to three treatment groups: treatment 1, a control (no additive); treatments 2 and 3, which consisted of two different additives (thiouracil and thyroxin respectively) to the rats drinking water \citep{box_1950}. Weight (grams) of the rats was measured at baseline (week 0) and at weeks 1, 2, 3, and 4. Data on three of the 10 rats from the thyroxin group were subsequently removed due to an accident at the beginning of the study. Figure~\ref{fig:4} shows estimated intercepts and slopes obtained from rat-specific LS regressions of the type
\begin{eqnarray*}
Y_{ij,k} = \beta_{0i,k} + \beta_{1i,k}x_{j} + \sigma_{i,k} \varepsilon_{ij,k}, & \; \varepsilon_{ij,k} \sim \mathcal{N}(0,1),
\end{eqnarray*}
where the response $Y_{ij,k}$ is weight measurement taken on rat $i = 1, \dots, M_{k}$ on occasion $j = 1, \ldots, 5$ conditional on treatment group $k = 1, 2, 3$, and $x_{j} = j - 1$. (Note that $M_{1} = M_{2} = 10$ and $M_{3} = 7$.) It is evident that the weight of rats treated with thiouracil grew slower than the controls', though at baseline the former tended to be heavier than the latter. In contrast, rats in the control and thyroxin groups had, on average, similar intercepts and slopes.

The Pearson's correlation coefficients of the estimated intercept-slope pairs $(\hat{\beta}_{0i,k},\hat{\beta}_{1i,k})$ gave $-0.26$ ($k = 1$), $-0.37$ ($k = 2$), and $-0.16$ ($k = 3$), suggesting a negative association between baseline weight and growth rate in all treatment groups. However, the direction of the association in treatment group 3 is unclear. Interestingly, the Kendall rank correlation coefficient in the thyroxin group indicated a weak positive association ($0.05$), while the Pearson's coefficient became strongly positive ($0.97$) after removing the two pairs with the largest slopes. Moreover, the distributions of intercepts and slopes showed the presence of skewness and bimodality. Therefore, some degree of robustness against departures from normality might be needed.

To model the heterogeneity within each treatment group, subject-specific random intercepts and slopes were included in the following four models: the NN model
\begin{align*}
Y_{ij,k} = \beta_{0,k} + \beta_{1,k}x_{j} + \nu^{(1)}_{1i,k} &+ \nu^{(2)}_{1i,k}x_{j} + \nu_{2ij,k}, \\ & \quad j = 1, \ldots, 5, \quad i = 1,\ldots,M_{k}, \quad k = 1,2,3,
\end{align*}
the NL model
\begin{align*}
Y_{ij,k} = \beta_{0,k} + \beta_{1,k}x_{j} + \nu^{(1)}_{1i,k} &+ \nu^{(2)}_{1i,k}x_{j} + \lambda_{2ij,k},\\ & \quad j = 1, \ldots, 5, \quad i = 1,\ldots,M_{k}, \quad k = 1,2,3,
\end{align*}
the LN model
\begin{align*}
Y_{ij,k} = \beta_{0,k} + \beta_{1,k}x_{j} + \lambda^{(1)}_{1i,k} &+ \lambda^{(2)}_{1i,k}x_{j} + \nu_{2ij,k},\\ & \quad j = 1, \ldots, 5, \quad i = 1,\ldots,M_{k}, \quad k = 1,2,3,
\end{align*}
and the LL model
\begin{align*}
Y_{ij,k} = \beta_{0,k} + \beta_{1,k}x_{j} + \lambda^{(1)}_{1i,k} &+ \lambda^{(2)}_{1i,k}x_{j} + \lambda_{2ij,k},\\ & \quad j = 1, \ldots, 5, \quad i = 1,\ldots,M_{k}, \quad k = 1,2,3,
\end{align*}
where I assumed $\left(\nu^{(1)}_{1i,k},\nu^{(2)}_{1i,k}\right) \sim \mathcal{N}_{2}(0, \Sigma_{1,k})$, $\nu_{2ij,k} \sim \mathcal{N}(0, \sigma_{2})$, $\left(\lambda^{(1)}_{1i,k},\lambda^{(2)}_{1i,k}\right) \sim \mathcal{L}_{2}(0,\Sigma_{1,k})$, and $\lambda_{2ij,k} \sim \mathcal{L}(0, \sigma_{2})$, and the $\Sigma_{1,k}$'s, $k = 1,2,3$, are $2 \times 2$ symmetric matrices,
\[
\Sigma_{1,k} = \left[\begin{array}{cc}
                \varsigma_{11,k} & \varsigma_{12,k} \\
                \varsigma_{12,k} & \varsigma_{22,k}
              \end{array}\right].
\]
Further, I assumed that the random effects are uncorrelated between treatment groups.

\begin{table}[t!]
\caption{Rats weight gain data. Estimates and standard errors (SE) of the fixed effects ($\beta$) from four models. The log-likelihood ($\ell$) is reported in brackets.}
\centering
\begin{tabular}{lrrrrrr}
  \toprule
 & $\beta_{0,1}$ & $\beta_{0,2}$ & $\beta_{0,3}$ & $\beta_{1,1}$ & $\beta_{1,2}$ & $\beta_{1,3}$ \\
  \hline
\multicolumn{7}{l}{\textit{Normal-Normal} ($\ell = -444.4$)}\\
Estimate & 52.880 & 57.700 & 52.086 & 26.480 & 17.050 & 27.143 \\
  SE &  2.349 & 2.058 & 1.578 & 1.177 & 0.879 & 1.928 \\
\multicolumn{7}{l}{\textit{Normal-Laplace} ($\ell = -448.4$)}\\
   Estimate & 52.934 & 57.568 & 52.928 & 26.383 & 17.208 & 26.791 \\
  SE & 2.427 & 2.204 & 1.519 & 1.208 & 0.928 & 2.146 \\
\multicolumn{7}{l}{\textit{Laplace-Normal} ($\ell = -551.6$)}\\
  Estimate & 53.069 & 58.392 & 51.104 & 25.620 & 16.794 & 26.665 \\
  SE & 1.992 & 1.972 & 1.817 & 0.885 & 0.814 & 1.910 \\
\multicolumn{7}{l}{\textit{Laplace-Laplace} ($\ell = -454.0$)}\\
  Estimate & 52.680 & 58.433 & 53.415 & 26.067 & 17.305 & 27.621 \\
  SE & 1.960 & 1.762 & 1.041 & 0.924 & 0.748 & 1.353 \\
   \hline
\end{tabular}\label{tab:3}
\end{table}

\begin{table}[ht!]
\caption{Rats weight gain data. Estimated correlation matrix of the random intercepts and slopes for each treatment group from three models. The log-likelihood ($\ell$) is reported in brackets.}
\centering
\begin{tabular}{lrrrrrr}
  \toprule
\multicolumn{7}{l}{\textit{Normal-Normal} ($\ell = -444.4$)}\\
  \hline
& \multicolumn{2}{c}{\textit{Treatment 1}} & \multicolumn{2}{c}{\textit{Treatment 2}} & \multicolumn{2}{c}{\textit{Treatment 3}}\\
 & Int. & Slope & Int. & Slope & Int. & Slope \\
Int. & 1.000 &  & 1.000 &  & 1.000 &  \\
Slope & $-$0.145 & 1.000 & $-$0.203 & 1.000 & 0.050 & 1.000 \\
  \hline
\multicolumn{7}{l}{\textit{Normal-Laplace} ($\ell = -448.4$)}\\
  \hline
& \multicolumn{2}{c}{\textit{Treatment 1}} & \multicolumn{2}{c}{\textit{Treatment 2}} & \multicolumn{2}{c}{\textit{Treatment 3}}\\
 & Int. & Slope & Int. & Slope & Int. & Slope \\
Int. & 1.000 &  & 1.000 &  & 1.000 &  \\
Slope & $-$0.076 & 1.000 & $-$0.133 & 1.000 & 0.634 & 1.000 \\
  \hline
\multicolumn{7}{l}{\textit{Laplace-Normal} ($\ell = -551.6$)}\\
  \hline
& \multicolumn{2}{c}{\textit{Treatment 1}} & \multicolumn{2}{c}{\textit{Treatment 2}} & \multicolumn{2}{c}{\textit{Treatment 3}}\\
 & Int. & Slope & Int. & Slope & Int. & Slope \\
Int. & 1.000 &  & 1.000 &  & 1.000 &  \\
Slope & $-$0.117 & 1.000 & $-$0.065 & 1.000 & 0.194 & 1.000 \\
  \hline
\multicolumn{7}{l}{\textit{Laplace-Laplace} ($\ell = -454.0$)}\\
  \hline
& \multicolumn{2}{c}{\textit{Treatment 1}} & \multicolumn{2}{c}{\textit{Treatment 2}} & \multicolumn{2}{c}{\textit{Treatment 3}}\\
 & Int. & Slope & Int. & Slope & Int. & Slope \\
Int. & 1.000 &  & 1.000 &  & 1.000 &  \\
Slope & 0.030 & 1.000 & $-$0.294 & 1.000 & 0.876 & 1.000 \\
   \hline
\end{tabular}\label{tab:4}
\end{table}

The NL, LN, and LL models were estimated using the EM algorithm as detailed in Appendix with a Monte Carlo size equal to $100$, fixed at each EM iteration, and a convergence tolerance of $5\cdot 10^{-4}$. The four models gave similar estimates of the fixed effects (Table \ref{tab:3}), although the trajectory in the thiouracil group resulting from the LN model tended to be less steep than the corresponding trajectory resulting from the other three models. However, this difference might be of little practical importance. In contrast, more substantial seemed to be the differences between the estimates of the correlation matrices $D_{1,k}^{-1}\Sigma_{1,k}D_{1,k}^{-1}$, where $D = \mathrm{diag}(\sqrt{\varsigma_{11,k}}, \sqrt{\varsigma_{22,k}})$, $k=1,2,3$ (Table \ref{tab:4}). It is interesting to note that there is disagreement on the magnitude and even direction of some of the estimates. Notably, $\hat{\varsigma}_{12,3}/(\hat{\varsigma}_{11,3} \cdot \hat{\varsigma}_{22,3})$ was smallest for the NN model but it was substantially larger for the NL and LL models. The best fit in terms of the log-likelihood was for the NN model, followed closely by the NL model. The LL model and, especially, the LN model gave smaller log-likelihoods.

\section{Final remarks}\label{sec:5}

In the words of \citet[][p.842]{wilson_1923} ``No phenomenon is better known perhaps, as a plain matter of fact, than that the frequencies which I actually meet in everyday work in economics, in biometrics, or in vital statistics, very frequently fail to conform at all closely to the so-called normal distribution''. Kotz and colleagues (\citeyear{kotz_2001}) echo Wilson's observations on the inadequacy of the normal distribution in many practical applications and give a systematic exposition of the Laplace distribution, an unjustifiably neglected error law which can be ``a natural and sometimes superior alternative to the normal law'' \citep[p.13]{kotz_2001}.

My proposed $2 \times 2$ convolution scheme brings together the normal and Laplace distributions showing that these models represent a \textit{family} of sensible alternatives as they introduce a varying degree of robustness in the modelling process. Estimation can be approached in different ways. The EM algorithm discussed in this paper takes advantage of the scale mixture representation of the Laplace distribution which provides the opportunity for computational simplification. In a simulation study with a moderate sample size (see Section~\ref{sec:A.7} in Appendix), this algorithm provided satisfactory results in terms of mean squared error for the NL and LL models. The estimation of the LN model needed a relatively larger number of Monte Carlo samples to achieve reasonable bias, though the results were never fully satisfactory in terms of efficiency. Finally, model selection has been left out of consideration, but further research on this topic is needed, especially at smaller sample sizes. An interesting starting point is offered by \cite{kundu_2005}.

To reiterate the main point of this study, these convolutions have a large number of potential applications and, as demonstrated using several examples, may provide valuable insight into different aspects of the analysis.

\section*{Acknowledgements}
This research has been supported by the National Institutes of Health -- National Institute of Child Health and Human Development (Grant Number: 1R03HD084807-01A1).

\appendix
\section*{Appendix}

\subsection{EM estimation}\label{sec:A.1}

Here, I discuss maximum likelihood inference for $\beta$, $\Sigma_1$, and $\sigma_2$ in normal-Laplace (NL), Laplace-normal (LN), and Laplace-Laplace (LL) models. In particular, I develop an estimation approach based on the scale mixture representation of the Laplace distribution. If $T \sim \mathcal{L}(0, \sigma)$ then $T \overset{d}{=} \sigma\sqrt{W}V$, where $W$ and $V$ are, respectively, independent standard exponential and normal variates. Similarly, if $T \sim \mathcal{L}_{q}(0,\Sigma_1)$, then $T \overset{d}{=} \sqrt{W}V$, where $W$ is, again, standard exponential and $V \sim \mathcal{N}_{q}(0, \Sigma_{1})$ \citep{kotz_2001}.

Let $Y_{i} = (Y_{i1}, Y_{i2}, \ldots, Y_{in_{i}})^{\top}$ be a multivariate $n_{i} \times 1$ random vector, and $x_{ij}$ and $z_{ij}$ be, respectively, $p \times 1$ and $q \times 1$ vectors of covariates for the $j$th observation, $j = 1, \ldots, n_{i}$, in cluster $i$, $i = 1, \ldots, M$. Also, let $X_{i}$ and $Z_{i}$ be, respectively, $n_{i} \times p$ and $n_{i} \times q$ design matrices for cluster $i$. I assume that the random effects have variance-covariance matrix $\Sigma_{1}$ for all $i = 1, \ldots, n$ and that the $Y_{i}$'s are independent of one another. The structure of $\Sigma_1$ is purposely left unspecified. The relative precision matrix $\sigma_{2}^{2}\Sigma_{1}^{-1}$ is parameterised in terms of an unrestricted $m$-dimensional vector, $1 \leq m \leq q(q+1)/2$, of non-redundant parameters $\xi$ \citep{pinheiro_bates}. The parameter to be estimated is then $\theta = \left(\beta^{\top},\xi^{\top},\sigma_{2}\right)^{\top}$ of dimension $(p + m + 1) \times 1$. The $n \times n$ identity matrix will be denoted by $I_{n}$.

\subsection{Normal-Laplace convolution}\label{sec:A.2}

Let $w_{i} = (w_{i1}, \ldots, w_{in_{i}})^{\top}$ be a $n_{i} \times 1$ vector of independent standard exponential variates and let $D_{i} = \mathrm{diag}(w_{i})$. The NL model can be written as
\begin{equation}\label{eq:A.1}
Y_{i} = X_{i}\beta + Z_{i}\nu_{1i} + D^{1/2}_{i}v_{i},
\end{equation}
where $\nu_{1i} \sim \mathcal{N}_{q}(0, \Sigma_{1})$ and $v_{i} \sim \mathcal{N}_{n_{i}}(0, \sigma^{2}_{2} I_{n_{i}})$. The model can be simplified by convolving $\nu_{1i}$ and $v_{i}$ conditional on $w_{i}$, i.e., by integrating out the random effects
\[
g\left(y_{i},w_{i}\right) = \int_{\mathbb{R}^q} g\left(y_{i},\nu_{1i}|w_{i}\right)h\left(w_{i}\right) \rds \nu_{1i} = g\left(y_{i}|w_{i}\right)h\left(w_{i}\right),
\]
where $y_{i}|w_{i} \sim \mathcal{N}_{n_{i}}\left(X_{i}\beta, \Omega_{i}\right)$, $\Omega_{i} = Z_{i}\Sigma_{1}Z_{i}^{\top} + \sigma^{2}_{2}D_{i}$, and $h\left(w_{i}\right) = \prod_{j = 1}^{n_{i}} \exp\left(-w_{ij}\right)$.

\subsection{Laplace-Normal convolution}\label{sec:A.3}

The LN model can be written as
\begin{equation}\label{eq:A.2}
Y_{i} = X_{i}\beta + \sqrt{w_{i}}Z_{i}v_{i} + \nu_{2i},
\end{equation}
where $w_{i}$ is a standard exponential variate, $v_{i} \sim \mathcal{N}_{q}(0, \Sigma_{1})$, and $\nu_{2i} \sim \mathcal{N}_{n_{i}}(0, \sigma^{2}_{2} I_{n_{i}})$. The normal component of the random effects can be integrated out as follows
\[
g\left(y_{i},w_{i}\right) = \int_{\mathbb{R}^q} g\left(y_{i},v_{i}|w_{i}\right)h\left(w_{i}\right) \rds v_{i} = g\left(y_{i}|w_{i}\right)h\left(w_{i}\right),
\]
where $y_{i}|w_{i} \sim \mathcal{N}_{n_{i}}\left(X_{i}\beta, \Omega_{i}\right)$, $\Omega_{i} = w_{i}Z_{i}\Sigma_{1}Z_{i}^{\top} + \sigma^{2}_{2} I_{n_{i}}$, and $h(w_{i}) = \exp\left(-w_{i}\right)$.

\subsection{Laplace-Laplace convolution}\label{sec:A.4}

Let $w_{i} = \left(w_{i,1},w_{i,2}^{\top}\right)^{\top}$ be a $(1 + n_{i}) \times 1$ vector of independent standard exponential variates, where $w_{i,2} = (w_{i1,2}, \ldots, w_{in_{i},2})^{\top}$, and let $D_{i} = \mathrm{diag}\left(w_{i,2}\right)$. The LL model can be written as
\begin{equation}\label{eq:A.3}
Y_{i} = X_{i}\beta + \sqrt{w_{i,1}}Z_{i}v_{1i} + D^{1/2}_{i}v_{2i},
\end{equation}
where $v_{1i} \sim \mathcal{N}_{q}(0, \Sigma_{1})$ and $v_{2i} \sim \mathcal{N}_{n_{i}}(0, \sigma^{2}_{2} I_{n_{i}})$. As before, the joint density can be simplified to
\[
g\left(y_{i},w_{i}\right) = \int_{\mathbb{R}^q} g\left(y_{i},v_{1i}|w_{i}\right)h\left(w_{i}\right) \rds v_{1i} = g\left(y_{i}|w_{i}\right)h\left(w_{i}\right),
\]
where $y_{i}|w_{i} \sim \mathcal{N}_{n_{i}}\left(X_{i}\beta, \Omega_{i}\right)$, $\Omega_{i} = w_{i,1}Z_{i}\Sigma_{1}Z_{i}^{\top} + \sigma^{2}_{2} D_{i}$, and $h\left(w_{i}\right) = \exp(-w_{i,1})\cdot \\ \prod_{j = 1}^{n_{i}} \exp\left(-w_{ij,2}\right)$.

\subsection{The algorithm}
\label{sec:A.5}

The joint density $g\left(y,w\right)$ could be further integrated to obtain $g\left(y\right) = \int g\left(y|w\right) \cdot h(w)\rds w$. Except for the normal-normal (NN) model, the form of the marginal likelihood of $Y_{i}$ does not seem to have an immediate known form. A numerical integration could have some appeal since this integral would reduce to a Gauss-Laguerre quadrature (h($w$) is standard exponential). However, since quadrature methods are notoriously inefficient if the dimension of the integral is large, I consider an alternative approach based on Monte Carlo EM (MCEM) estimation. In this case, the unobservable variable $w$ is sampled from the conditional density $g(w|y)$. While the Monte Carlo sample size does not depend as much on dimensionality as quadrature methods do, convergence can be slower for MCEM than for quadrature-based methods \citep{mclachlan_2008}.

The $i$th contribution to the complete data log-likelihood for the models (\ref{eq:A.1})-(\ref{eq:A.3}) is given by
\begin{equation}\label{eq:A.4}
\ell\left(\theta; y_{i},w_{i}\right) = \log g\left(y_{i}|w_{i}\right) + \log h(w_{i}).
\end{equation}
Note that $h(w_{i})$ does not depend on $\theta$. The EM approach alternates between an
\begin{itemize}
\item[(i)] expectation step (E-step) $Q_{i}(\theta|\theta^{(t)}) = \expect_{w|y,\theta^{(t)}}\left\{\ell\left(\theta; y_{i},w_{i}\right) \right\}$, $i = 1, \ldots, M$; and a
\item[(ii)] maximisation step (M-step) $\theta^{(t+1)} = \underset{\theta}{\operatorname{arg\,max}} \ \sum_{i} Q_{i}(\theta|\theta^{(t)})$,
\end{itemize}
where $\theta^{(t)}$ is the estimate of the parameter after $t$ cycles. The expectation in step (i) is taken with respect to $h\left(w_{i}|y_{i},\theta^{(t)}\right) \propto g\left(y_{i}|w_{i},\theta^{(t)}\right)h(w_{i})$, that is, the distribution of the unobservable data $w_{i}$ conditional on the observed data $y_{i}$ and the current estimate of $\theta$. Given that the latter density does not have an immediate known form, I consider a Monte Carlo approach and use the following numerical approximation
\begin{equation}\label{eq:A.5}
\tilde{Q}_{i}(\theta|\theta^{(t)}) = \dfrac{1}{K}\sum_{k=1}^{K}\left\{\ell\left(\theta; y_{i},w^{(t)}_{ik}\right) \right\},
\end{equation}
where $w^{(t)}_{ik}$ is a vector of appropriate dimensions sampled from $h\left(w_{i}|y_{i},\theta^{(t)}\right)$ at iteration $t$. The number of samples $K$ can be fixed at the same value for all iterations or may vary with $t$. The approximate complete data log-likelihood for all clusters (Q-function), averaged over $w|y$, is given by
\begin{align}\label{eq:A.6}
\tilde{Q} (\theta|\theta^{(t)})\equiv \sum_{i=1}^{M}\tilde{Q}_{i}(\theta|\theta^{(t)}) = & \dfrac{1}{K}\sum_{k=1}^{K}\sum_{i=1}^{M}- \frac{n_{i}}{2}\log(2\pi) -\dfrac{1}{2}\log |\Omega_{ik}|\\
\nonumber & - \frac{1}{2}e_{i}^{\top}{\Omega_{ik}}^{-1}e_{i} + \log h\left(w^{(t)}_{ik}\right),
\end{align}
where $e_{i}=y_{i}-X_{i}\beta$, $\Omega_{ik} = \sigma_{2}^{2} \Psi_{ik}$,
\begin{equation*}
\Psi_{ik} =
\begin{cases}
Z_{i}\dot{\Sigma}_{1}Z_{i}^{\top} + D_{ik} & \text{with $D_{ik} = \mathrm{diag}\left(w^{(t)}_{ik}\right)$ for the NL model,}\\[5pt]
w^{(t)}_{ik}Z_{i}\dot{\Sigma}_{1}Z_{i}^{\top} + I_{n_{i}} & \text{for the LN model,}\\[5pt]
w^{(t)}_{ik,1}Z_{i}\dot{\Sigma}_{1}Z_{i}^{\top} + D_{ik} & \text{with $D_{ik} = \mathrm{diag}\left(w^{(t)}_{ik,2}\right)$ for the LL model,}\\
\end{cases}
\end{equation*}
and $\dot{\Sigma}_{1}= \sigma_{2}^{-2}\Sigma_{1}$ is the scaled variance-covariance matrix of the random effects. Note that all the information given by $\theta^{(t)}$ is contained in $\Omega_{ik}$ which depends on $w^{(t)}_{ik}$ (the superscript $(t)$ has been dropped from $\Omega_{ik}$, $\Psi_{ik}$, and $D_{ik}$ to ease notation). Furthermore, the parameter $\xi$ is defined to be the vector of non-zero elements of the upper triangle of the matrix logarithm of $U$, where $U$ is the $q \times q$ matrix obtained from the Cholesky decomposition $\dot{\Sigma}^{-1}_{1} = U^{\top}U$ \citep{pinheiro_bates}.

The Q-function (\ref{eq:A.6}) can be easily maximised with respect to $\beta$, $\xi$, and $\sigma_{2}$ using standard (restricted) maximum likelihood formulas for linear mixed models \citep{pinheiro_bates,demidenko_2013}. Indeed, the derivative of (\ref{eq:A.6}) with respect to $\theta$ has the familiar form
\begin{equation}\label{eq:A.7}
\tilde{Q}_{\ast}(\theta|\theta^{(t)}) =
\left(\begin{array}{c}
\dfrac{1}{K}\sum_{k=1}^{K}\sum_{i=1}^{M} \sigma_{2}^{-2} X_{i} \Psi_{ik}^{-1}e_{i}\\[10pt]
-\dfrac{1}{2K}\sum_{i=1}^{M}\sum_{k=1}^{K} Z_{i}^{\top}\Psi_{ik}^{-1}Z_{i} - \sigma_{2}^{-2}Z_{i}^{\top}\Psi_{ik}^{-1}e_{i}e_{i}^{\top}\Psi_{ik}^{-1}Z_{i}\\[10pt]
-\dfrac{1}{2}N\sigma_{2}^{-2}+\dfrac{1}{2K}\sigma_{2}^{-4}\sum_{k=1}^{K}\sum_{i=1}^{M}e_{i}^{\top}\Psi_{ik}^{-1}e_{i}
\end{array}\right),
\end{equation}
where $N = \sum_{i}^{M}n_{i}$. Since the system of equations $\tilde{Q}_{\ast}(\theta|\theta^{(t)}) = 0$ does not have a simultaneous closed-form solution, we must resort to an iterative algorithm (e.g., Newton--Raphson). Note, however, that for fixed $\Psi_{ik}$ at iteration $t$, the Q-function is maximised by
\[
\hat{\beta} = \left(\sum_{i=1}^{M}X_{i}^{\top}\Psi_{ik}^{-1}X_{i}\right)^{-1}\left(\sum_{i=1}^{M}X_{i}^{\top}\Psi_{ik}^{-1}y_{i}\right).
\]
Thus, for the NL, LN, and LL models, the EM estimate $\hat{\beta}$ can be seen as the solution of the generalised least squares (GLS) with weights that depend on the sampled values $w_{ik}^{(t)}$. The variance-covariance of $\hat{\beta}$,
\[
\mathrm{cov}\left(\hat{\beta}\right) = \sigma^{2}_{2}\left(\sum_{i=1}^{M} X_{i}^{\top}\Psi_{ik}^{-1}X_{i}\right)^{-1},
\]
is a by-product of fitting routines from commonly available software.

Similarly, for fixed $\Psi_{ik}$ at iteration $t$, the GLS estimate of $\sigma_{2}^{2}$ is
\[
\hat{\sigma}_{2}^{2} = \left(\sum_{i=1}^{M}y_{i}^{\top}\Psi_{ik}^{-1}y_{i}\right) - \left(\sum_{i=1}^{M}X_{i}^{\top}\Psi_{ik}^{-1}y_{i}\right)^{\top}\left(\sum_{i=1}^{M}X_{i}^{\top}\Psi_{ik}^{-1}X_{i}\right)^{-1}
\left(\sum_{i=1}^{M}X_{i}^{\top}\Psi_{ik}^{-1}y_{i}\right).
\]

The E-step is updated with $\theta^{(t+1)}$ and the algorithm stops when\\$\Delta_{h:t,t+1} \left\{\tilde{Q} (\theta|\theta^{(h)})\right\} < \delta$ or $\Delta_{h:t,t+1} \left\{\theta_{l}^{(h)}\right\} < \delta$, $l = 1,\ldots,p + m + 1$, where $\Delta_{h:t,t+1}\left\{u^{(h)}\right\}$ is the (absolute or relative) change in $u$ between iterations $t$ and $t+1$, and $\delta$ is an appropriately small constant. The starting values $\theta^{(0)}$ can be obtained from an LME model.

Finally, standard errors for $\hat{\theta}$ can be computed using the methods described in \cite{mclachlan_2008}. See also the application of Rubin's rules for multiple imputation to Monte Carlo EM samples \citep{goetghebeur_2000,geraci_farcomeni}.

\subsection{Residual heteroscedasticity and correlation}
\label{sec:A.6}

In the previous sections, I assumed that the within-group errors are independent with common scale parameter $\sigma_{2}$. Using the scale mixture representation, it is immediate to extend the NL, LN, and LL models to the case of heteroscedastic and correlated errors. In particular, let's assume $\lambda_{2i} \sim \mathcal{L}_{n_{i}}(0,\Sigma_{2i})$ for the NL and LL convolutions, and $\nu_{2i} \sim \mathcal{N}_{n_{i}}(0, \Sigma_{2i})$ for the LN convolution, with general $\Sigma_{2i}$, $i = 1,\dots,M$. Then the variance-covariance matrix in (\ref{eq:A.6}) can be written as
\begin{equation*}
\Omega_{ik} =
\begin{cases}
Z_{i}\Sigma_{1}Z_{i}^{\top} + w^{(t)}_{ik}\Sigma_{2i} & \text{for the NL model,}\\[5pt]
w^{(t)}_{ik}Z_{i}\Sigma_{1}Z_{i}^{\top} + \Sigma_{2i} & \text{for the LN model,}\\[5pt]
w^{(t)}_{ik,1}Z_{i}\Sigma_{1}Z_{i}^{\top} + w^{(t)}_{ik,2}\Sigma_{2i} & \text{for the LL model.}\\
\end{cases}
\end{equation*}

\subsection{Monte Carlo}
\label{sec:A.7}

In this section, I report on the results of a small simulation study. The purpose was to investigate the bias, variance, and mean squared error (MSE) of $\hat{\beta}$ and $\hat{\xi}$ for the NN, NL, LN, and LL models when data were generated according to the following four scenarios:
\begin{enumerate}
\item $Y_{ij} = x_{ij}^{\top}\beta + z_{ij}^{\top}\nu_{1i} + \nu_{2ij}$,
\item $Y_{ij} = x_{ij}^{\top}\beta + z_{ij}^{\top}\nu_{1i} + \lambda_{2ij}$,
\item $Y_{ij} = x_{ij}^{\top}\beta + z_{ij}^{\top}\lambda_{1i} + \nu_{2ij}$,
\item $Y_{ij} = x_{ij}^{\top}\beta + z_{ij}^{\top}\lambda_{1i} + \lambda_{2ij}$,
\end{enumerate}
where $\beta = (\beta_{0}, \beta_{1})^{\top} = (1, 2)^{\top}$, $x_{ij} = (1, x_{1ij})^{\top}$, $z_{ij} = x_{ij}$, with $x_{1ij} = \gamma_{i} + \zeta_{ij}$, $\gamma_{i}\sim \mathcal{N}(0,1)$, and $\zeta_{ij}\sim \mathcal{N}(0,1)$. The random effects were sampled from multivariate normal ($\nu_{1}$) or Laplace ($\lambda_{1}$) distributions with variance-covariance
\[
\Sigma_{1} = \left[\begin{array}{cc}
                3 & 1 \\
                1 & 2
              \end{array}\right],
\]
while the errors were drawn from normal ($\nu_{2}$) or Laplace ($\lambda_{2}$) distributions with scale $\sigma_2 = 2$, independently. The unrestricted parameter for $\sigma_{2}^{2}\Sigma_{1}^{-1}$ is given by $\xi = (\xi_{1}, \xi_{2}, \xi_{3})^{\top} = (-0.183, 0.215, -0.398)^{\top}$.

A balanced design with $n = 5$ repeated measurements per cluster and $M = 100$ clusters was used. For each scenario, 100 datasets were replicated. NN models were fitted using MLE routines from the \texttt{nlme} package \citep{pinheiro_2014}. The NL, LN, and LL models were fitted using the EM algorithm discussed above. In particular, Monte Carlo samples for the E-step were drawn using an adaptive rejection Metropolis sampler \citep{gilks} as implemented in the package \texttt{HI} \citep{petris}. The number of samples was set to increase at each EM iteration as a multiple of 20, capped at 500, thus $K = \min\{20\cdot t,500\}$, $t = 1, 2, \ldots$. The Q-function (\ref{eq:A.6}) was maximised using the MLE equations for NN models \citep{demidenko_2013}. The convergence criterion was defined as $\Delta_{h:t,t+1} \left\{\tilde{Q} (\theta|\theta^{(h)})\right\} < 0.001$ and the maximum number of EM iterations was set to 100.

The results of the simulation study are reported in Tables~\ref{tab:5}-\ref{tab:7}. The NL and LL showed some advantages in terms of bias as compared to the NN model in all considered scenarios, including when the data were generated from a NN model. However, in the latter case the lower bias was more than compensated by a larger variability which made the MSE for NL and LL models up to about $44\%$ larger than that for the NN model. In contrast, the NN model was less competitive than the NL and LL models when data were generated according to these two scenarios, with losses up to about $40\%$ in terms of MSE.

The LN model's performance was somewhat poor, even when the data were generated from a LN model. In a separate analysis using the same data (results not shown), the NL models were re-estimated with the number $K$ of Monte Carlo samples fixed at 500 at all iterations. The relative bias decreased to values below or near 1 for both $\hat{\beta}$ and $\hat{\xi}$, whereas the relative MSE was still above 1.

The average estimation times (standard deviation) for the NL, LN, and LL models were, respectively, 6.6 (10.0), 17.7 (25.3), and 9.8 (11.0) minutes on a 64-bit operating system machine with 16 Gb of RAM and quad-core processor at 3.60 GHz. The average number of iterations to convergence for all these three models was 14 (standard deviation 12).

\begin{table}[h!]
\caption{The estimated bias of $\hat{\beta}$ and $\hat{\xi}$ for the normal-normal (NN) model is reported in brackets. The bias for the normal-Laplace (NL), Laplace-normal (LN), and Laplace-Laplace (LL) models is relative to the NN model.}
\centering
\begin{tabular}{lrrrrr}
\toprule
 & $\hat{\beta}_{0}$ & $\hat{\beta}_{1}$ & $\hat{\xi}_{1}$ & $\hat{\xi}_{2}$ & $\hat{\xi}_{3}$\\
\hline
\multicolumn{6}{l}{\textit{Scenario 1: NN data}}\\
\hline
NN & ($-$0.012) & (0.006) & ($-$0.036) & (0.007) & ($-$0.035) \\
NL & 0.324 & 0.254 & 1.648 & $-$0.418 & 1.977 \\
LN & 0.088 & 1.424 & $-$1.823 & 8.010 & $-$2.997 \\
LL & 0.840 & $-$0.369 & $-$0.402 & 0.784 & $-$0.001 \\
\hline
\multicolumn{6}{l}{\textit{Scenario 2: NL data}}\\
\hline
NN & (0.002) & ($-$0.016) & ($-$0.021) & (0.014) & ($-$0.062) \\
NL & $-$1.301 & 0.503 & 0.970 & 0.848 & 0.881 \\
\hline
\multicolumn{6}{l}{\textit{Scenario 3: LN data}}\\
\hline
NN & ($-$0.005) & (0.020) & ($-$0.072) & (0.044) & ($-$0.042) \\
LN & 1.731 & 1.232 & $-$0.243 & 2.363 & $-$1.822 \\
\hline
\multicolumn{6}{l}{\textit{Scenario 4: LL data}}\\
\hline
NN & ($-$0.006) & (0.012) & ($-$0.029) & (0.023) & ($-$0.079) \\
LL & 0.716 & 0.483 & $-$0.504 & 0.829 & 0.345 \\
\hline
\end{tabular}\label{tab:5}
\end{table}

\begin{table}[h!]
\caption{The estimated variance of $\hat{\beta}$ and $\hat{\xi}$ for the normal-normal (NN) model is reported in brackets. The variance for the normal-Laplace (NL), Laplace-normal (LN), and Laplace-Laplace (LL) models is relative to the NN model.}
\centering
\begin{tabular}{lrrrrr}
\toprule
 & $\hat{\beta}_{0}$ & $\hat{\beta}_{1}$ & $\hat{\xi}_{1}$ & $\hat{\xi}_{2}$ & $\hat{\xi}_{3}$\\
\hline
\multicolumn{6}{l}{\textit{Scenario 1: NN data}}\\
\hline
NN & (0.040) & (0.028) & (0.012) & (0.005) & (0.014) \\
NL & 1.123 & 1.081 & 1.021 & 0.994 & 1.259 \\
LN & 1.362 & 1.842 & 2.207 & 2.709 & 1.665\\
LL & 1.287 & 1.441 & 1.174 & 1.246 & 1.087 \\
\hline
\multicolumn{6}{l}{\textit{Scenario 2: NL data}}\\
\hline
NN & (0.048) & (0.029) & (0.018) & (0.007) & (0.016) \\
NL & 0.894 & 1.042 & 0.925 & 0.904 & 0.843 \\
\hline
\multicolumn{6}{l}{\textit{Scenario 3: LN data}}\\
\hline
NN & (0.034) & (0.037) & (0.039) & (0.015) & (0.040) \\
LN & 1.126 & 1.146 & 2.020 & 2.211 & 1.691 \\
\hline
\multicolumn{6}{l}{\textit{Scenario 4: LL data}}\\
\hline
NN & (0.043) & (0.024) & (0.026) & (0.011) & (0.026) \\
LL & 0.744 & 0.862 & 0.813 & 0.598 & 0.997 \\
\hline
\end{tabular}\label{tab:6}
\end{table}

\begin{table}[h!]
\caption{The estimated mean squared error (MSE) of $\hat{\beta}$ and $\hat{\xi}$ for the normal-normal (NN) model is reported in brackets. The MSE for the normal-Laplace (NL), Laplace-normal (LN), and Laplace-Laplace (LL) models is relative to the NN model.}
\centering
\begin{tabular}{lrrrrr}
\toprule
 & $\hat{\beta}_{0}$ & $\hat{\beta}_{1}$ & $\hat{\xi}_{1}$ & $\hat{\xi}_{2}$ & $\hat{\xi}_{3}$\\
\hline
\multicolumn{6}{l}{\textit{Scenario 1: NN data}}\\
\hline
NN & (0.040) & (0.028) & (0.013) & (0.005) & (0.016) \\
NL & 1.119 & 1.080 & 1.190 & 0.987 & 1.471 \\
LN & 1.357 & 1.842 & 2.319 & 3.242 & 2.248 \\
LL & 1.285 & 1.440 & 1.073 & 1.240 & 1.001 \\
\hline
\multicolumn{6}{l}{\textit{Scenario 2: NL data}}\\
\hline
NN & (0.048) & (0.029) & (0.018) & (0.007) & (0.020) \\
NL & 0.894 & 1.035 & 0.925 & 0.898 & 0.830 \\
\hline
\multicolumn{6}{l}{\textit{Scenario 3: LN data}}\\
\hline
NN & (0.034) & (0.038) & (0.044) & (0.017) & (0.042) \\
LN & 1.127 & 1.150 & 1.787 & 2.598 & 1.759 \\
\hline
\multicolumn{6}{l}{\textit{Scenario 4: LL data}}\\
\hline
NN & (0.043) & (0.025) & (0.027) & (0.012) & (0.033) \\
LL & 0.744 & 0.859 & 0.796 & 0.602 & 0.830 \\
\hline
\end{tabular}\label{tab:7}
\end{table}

\end{document}